\documentclass[twocolumn,superscriptaddress]{revtex4-2}
\usepackage{xcolor}
\usepackage{graphicx}
\usepackage{dcolumn}
\usepackage{bm}
\usepackage[colorlinks=true, a4paper=true, pdfstartview=FitV,
linkcolor=blue, citecolor=blue, urlcolor=blue]{hyperref}

\newcommand{\be}{\begin{equation}}
\newcommand{\ee}{\end{equation}}
\newcommand{\bea}{\begin{eqnarray}}
\newcommand{\eea}{\end{eqnarray}}
\newcommand{\bean}{\begin{eqnarray*}}
\newcommand{\eean}{\end{eqnarray*}}

\begin{document}
\definecolor{mygreen}{HTML}{006E28}
\newcommand{\kasia}[1]{{\color{mygreen}\textbf{?KS:}  #1}}
\newcommand{\trom}[1]{{\color{red}\textbf{?TR:} \color{red} #1}}
\newcommand{\filip}[1]{\textbf{?FB:} {\color{blue} #1}}
\newcommand{\aw}[1]{{\color{magenta}\textbf{?AW:}  #1}}
%\preprint{APS/123-QED}

\title{Amplitude modulations and resonant decay of excited oscillons}
\author{F. Blaschke}

\affiliation{Research Center for Theoretical Physics and Astrophysics, Institute of Physics, Silesian University in Opava, Bezru\v{c}ovo n\'{a}m\v{e}st\'{\i}~1150/13, 746~01 Opava, Czech Republic
}
\affiliation{Institute of Experimental and Applied Physics, Czech Technical University in Prague, Husova~240/5, 110~00 Prague~1, Czech Republic}

\author{T. Roma\'{n}czukiewicz}
%\email[]{tomasz.romanczukiewicz@uj.edu.pl}
\affiliation{
 Institute of Theoretical Physics,  Jagiellonian University, Lojasiewicza 11, 30-348 Krak\'{o}w, Poland
}

\author{K. S\l{}awi\'{n}ska}
%\email[]{tomasz.romanczukiewicz@uj.edu.pl}
\affiliation{
 Institute of Theoretical Physics,  Jagiellonian University, Lojasiewicza 11, 30-348 Krak\'{o}w, Poland
}

\author{A. Wereszczy\'{n}ski}
%\email[]{andrzej.wereszczynski@uj.edu.pl}

\affiliation{
 Institute of Theoretical Physics,  Jagiellonian University, Lojasiewicza 11, 30-348 Krak\'{o}w, Poland
}
 \affiliation{Department of Applied Mathematics, University of Salamanca, Casas del Parque 2, 37008 - Salamanca, Spain
}

\affiliation{International Institute for Sustainability with Knotted Chiral Meta Matter (WPI-SKCM2), Hiroshima University, Higashi-Hiroshima, Hiroshima 739-8526, Japan}

\begin{abstract}
We show that the decay of strongly excited oscillons in a single vacuum model reveals a chaotic, fractal-like pattern very much like one found in kink-antikink collision in the $\phi^4$ model. This structure can be attributed to the resonant energy transfer mechanism triggered by the modulations of amplitudes of {\it constituent oscillons} which form the excited oscillon. We also find evidence that such modulations arise as a motion of two quasi-breathers inside the constituent oscillon. 
\end{abstract}

\maketitle

%\tableofcontents
%%%%%%%%%%%%%%%%%%%%%%%%%%%
\section{\label{sec:motiv}Motivation}
%%%%%%%%%%%%%%%%%%%%%%%%%%%
Oscillons \cite{G} are localized, quasi-periodic field excitations with a surprisingly long lifetime \cite{BM, GS, I, Z}. They have a broad range of applications, especially in astrophysical and cosmological contexts \cite{CGM, G-cosm, Amin-1, A}, where they are expected to be a reminiscence of phase transitions in the inflation field \cite{FMPW, AE, LT}. They also exist in the standard model of particle physics \cite{G-weak}. 

On the contrary to other non-perturbative solutions, there are no obvious arguments explaining their existence. There are no topological or nontopological reasons (conserved charged) that would prevent them from decay. In fact, despite a lot of work, e.g. \cite{G, Fod}, oscillons are still rather mysterious objects, whose origin, as well as properties, are awaiting a full explanation. See e.g., surprising discovery of oscillons in massless models \cite{DRSW}.

Even as basic feature as the identification of degrees of freedom (DoF), which could explain a characteristic double-frequency behavior (modulations of the amplitude), is still a challenge. Recently, an intriguing solution to this problem has been proposed, in which the dynamics of an oscillon, and in particular its double oscillations, were captured by modes of a sphaleron from which the oscillon is created \cite{MR}. Although this picture worked well for the analyzed $\phi^3$ oscillon, it does not seem to be a general mechanism of the modulations. There are models with amplitude-modulated oscillons that emerge from sphalerons with too few internal modes or where sphaleron simply does not exist \cite{OQRW}. 
 
In the present paper, we show that decay of strongly excited oscillon reveals a practically identical fractal structure as the one observed in kink-antikink collisions, e.g. in the $\phi^4$ model \cite{sug, CSW, MORW}. This happens even though the theory does not support any static, particle-like objects like kinks or sphalerons. The oscillons are the only localized excitations. Thus, the chaotic structure, usually associated with the resonant transfer phenomenon involving kinetic and internal degrees of freedom (DoF) of kinks or sphalerons, must originate entirely from oscillons themselves. 

The identification of the pertinent, oscillon-based DoF entering the resonant energy transfer is based on the observation that strongly excited oscillons are, in fact, multi-oscillon states, where {\it constituent}, weakly excited (with modulations) oscillons perform several bounces forming an oscillating quasi-bound state. 
In such a bound state kinetic motion of the constituent oscillons can be compensated by an attractive interaction arising from the amplitude modulation. 

Importantly, this composite picture reaches even deeper. Namely, the constituent oscillons can also be viewed as bound states of two {\it fundamental} (non-modulated, i.e. single frequency) oscillons identified with sine-Gordon breathers. Their mutual motion gives rise to the modulation of the amplitude.  
%%%%%%%%%%%%%%%%%%%%%%%%%%%
\section{\label{sec:model} Quasi-quadratic theory}
%%%%%%%%%%%%%%%%%%%%%%%%%%%
\begin{figure}
\center
\includegraphics[width=1.0\columnwidth]{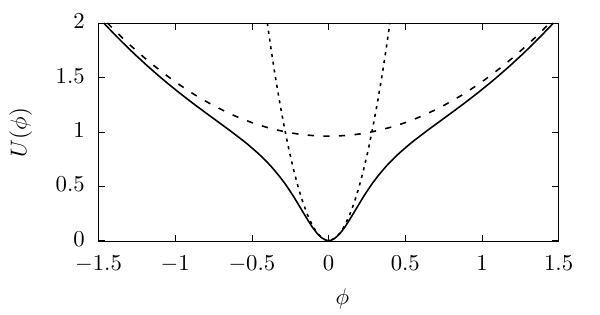}
\caption{The potential $U(\phi)$ (black) which interpolates between two quadratic potentials: $\frac{m^2}{2}\phi^2$ (dotted) and $\frac{1}{2}\phi^2+1-m^{-2}$ (dashed). Here  $m=5$.}
\label{plot-pot}
\end{figure}
To study the dynamics of oscillons we use a real scalar field theory with a simple single vacuum potential $U(\phi)$ which interpolates between two quadratic regimes. Specifically, 
\be
L=  \int_{-\infty}^\infty \left[ \frac{1}{2}\phi_t^2 -\frac{1}{2}\phi_x^2  -U(\phi) \right] dx, \label{Lag}
\ee 
where
\be
U(\phi)=\frac{1}{2}m^2 \phi^2 g(\phi) + \frac{1}{2}\phi^2 (1-g(\phi)) \label{pot}
\ee
and
\be
g(\phi)=\left(1 + m^2\frac{\phi^2}{2} \right)^{-1}
\ee
is a function interpolating between 1 and 0 when $\phi$ changes from the vacuum value, $\phi=0$, to infinity. $m$ is a mass of infinitesimally small perturbations over the vacuum. We take $m=5$ throughout. 

In this simple theory, there are no kinks or sphalerons and oscillons are the only localized excitations. As a consequence, the dynamics of oscillons is not obscured by complicated interactions with other localized (solitonic) degrees of freedom.

% In other words, the only role of nonlinearities is to smoothly join these two sectors. Oscillons are the only localized excitations. There are no kink or sphaleron solutions. As a consequence, the dynamics of oscillons is not obscured by complicated interactions with other localized (solitonic) degrees of freedom. We take $m=5$ throughout.

%%%%%%%%%%%%%%%%%%%%%%%%%%%
\section{\label{sec:decay}Resonant decay of oscillons}
%%%%%%%%%%%%%%%%%%%%%%%%%%%
\begin{figure}
\center
\includegraphics[width=1.0\columnwidth]{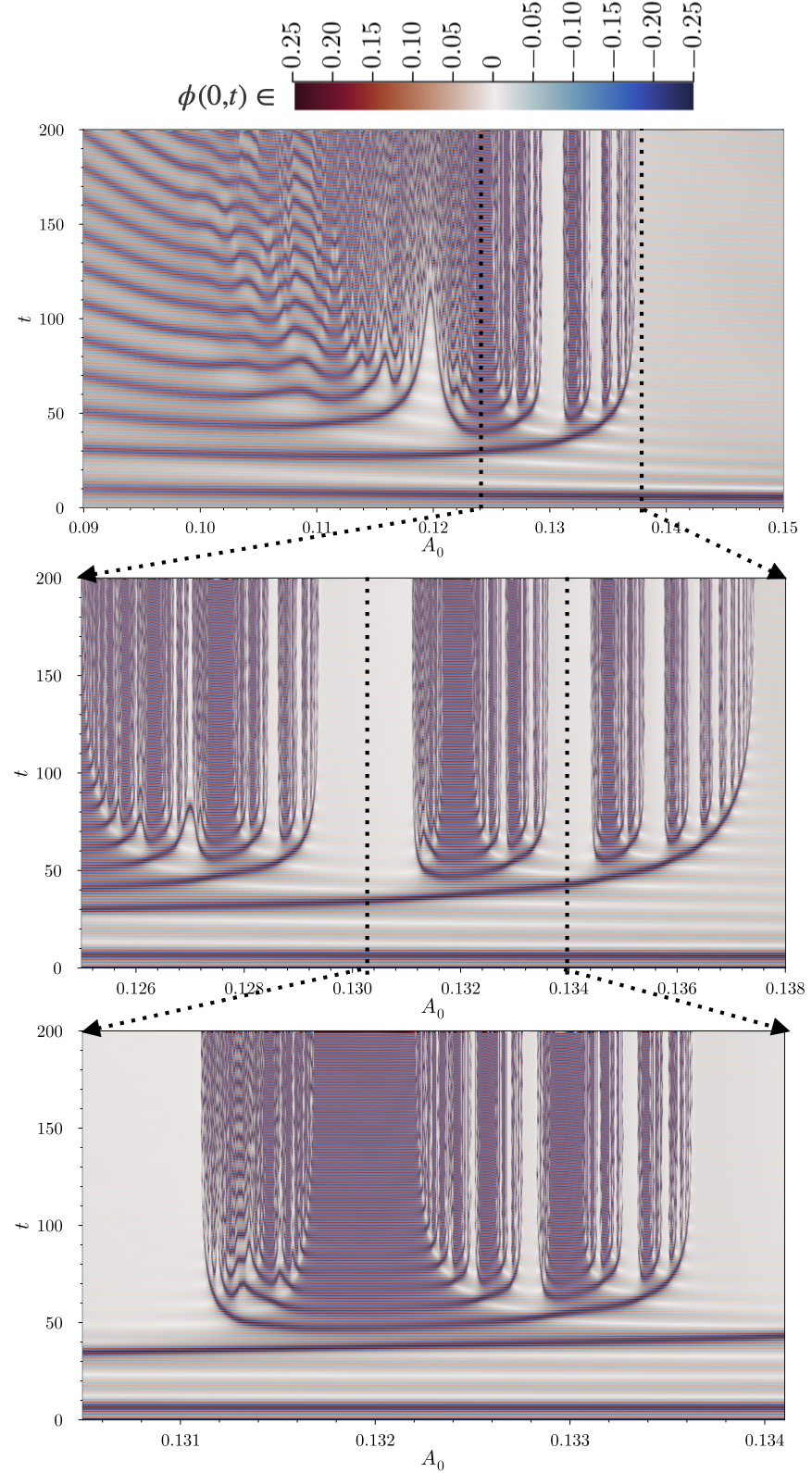}%{combinedmaps.pdf}
\caption{Time evolution of the value of the field at the origin $\phi(0,t)$ for different initial amplitude $A_0$. Here, $\sigma=10$.}
\label{plot-fractal}
\end{figure}
In our numerical experiments, we investigated oscillons produced from a Gaussian initial data 
\be
\phi(x,0)=A_0\, e^{-x^2/\sigma},
\ee
where $A_0$ and $\sigma$ are an amplitude and a size (squared) of the perturbation. Depending on the value of these parameters we find more or less excited oscillon which reveals various scenarios of evolution. We identify {\it unexcited} oscillon with a single frequency object. 

In Fig. \ref{plot-fractal} we plot the time dependence of the field at the origin, $\phi(0,t)$ as a function of the initial amplitude $A_0$. We chose $\sigma=10$. A detailed analysis of the evolution of different initial data is discussed in the supplementary materials. Amazingly, we find an almost identical fractal pattern as in the final state formation in kink-antikink collisions, e.g. in $\phi^4$ model \cite{sug, CSW}. In this case, such a structure is triggered by the famous {\it resonant energy transfer} between kinetic and internal degrees of freedom of the colliding solitons \cite{sug, CSW, MORW}. Here, however, there are no solitonic solutions, and therefore the resonant energy transfer must arise from DoF provided exclusively by the oscillons. 

\begin{figure*}
%\includegraphics[width=1.90\columnwidth]
%{combinedexamples3.pdf}
%{example3_m5_A0.13_sigma10_en.pdf}
%\includegraphics[width=0.68\columnwidth]{example3_m5_A0.134_sigma10_en.pdf}
%\includegraphics[width=0.68\columnwidth]{example3_m5_A0.1355_sigma10_en.pdf}
%\includegraphics[width=0.68\columnwidth]{example3_m5_A0.1364_sigma10_en.pdf}
%\includegraphics[width=0.68\columnwidth]{example3_m5_A0.1367_sigma10_en.pdf}
%\includegraphics[width=0.68\columnwidth]
%{example3_m5_A0.13694_sigma10_en.pdf}
%$0.1355$, $0.1364$, $0.1367$, $0.13694$}
%\label{2-b}
%\vspace*{0.4cm}
%\hspace*{-0.5cm} 
\includegraphics[width=1.80\columnwidth]
{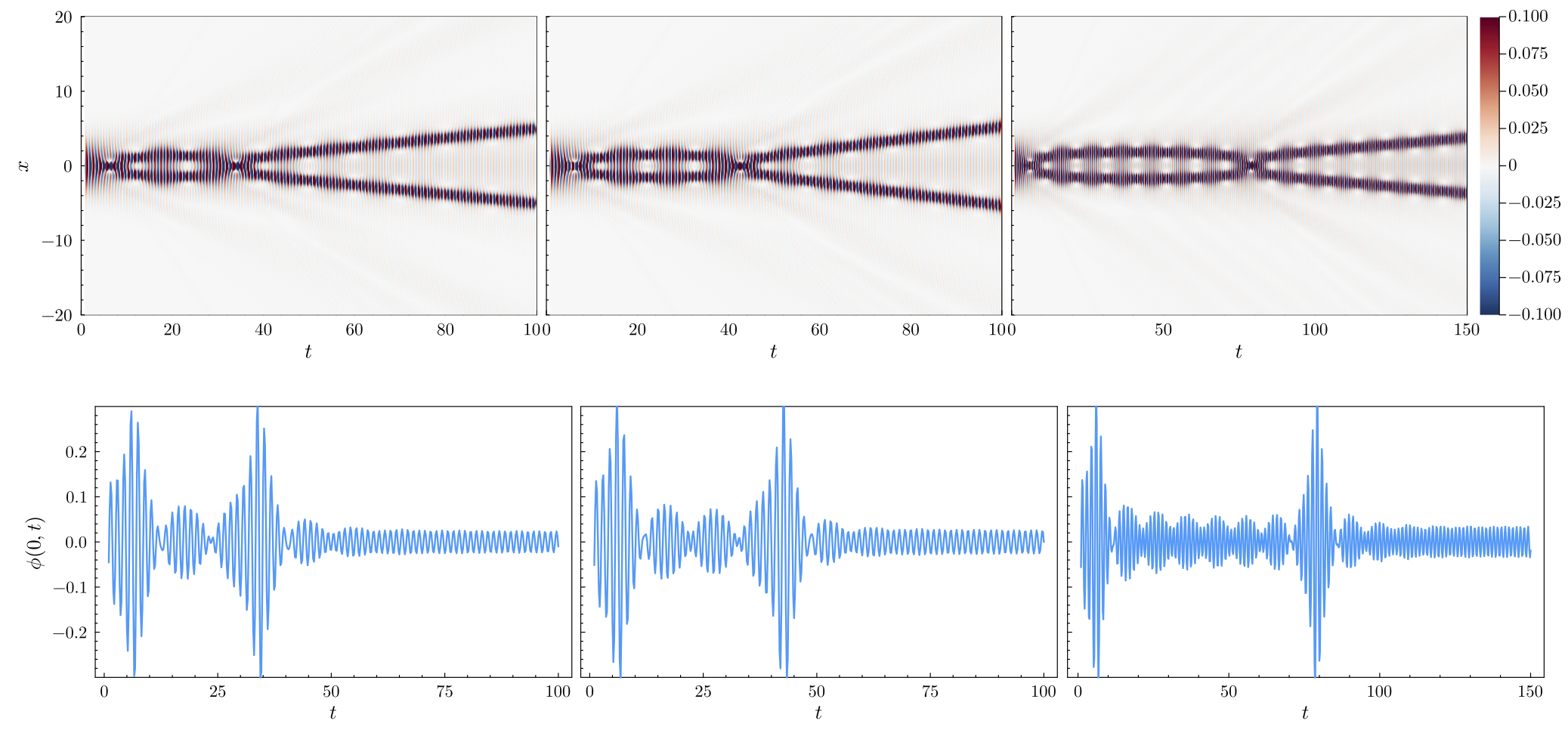}
%{field_m5_A0.13_sigma10.pdf}
%\includegraphics[width=0.68\columnwidth]{field_m5_A0.134_sigma10.pdf}
%\includegraphics[width=0.68\columnwidth]{field_m5_A0.13694_sigma10.pdf}
\caption{2-bounces of two oscillons with increasing number of modulations of the amplitude. Here $A_0=0.13$, $0.134$,  $0.13694$. Upper: $\phi(x,t)$. Lower $\phi(0,t).$}
%\caption{Value of the field at the origin for $A_0=0.13$, $0.134$, $0.13694$.}
\label{2-b}
\end{figure*}

To better understand this plot and identify DoF responsible for the fractal structure we show the field evolution in the full space for particular values of the amplitude $A_0$, see Fig. \ref{2-b}.  In the upper panel, we show examples of dynamics where the initial data forms an excited oscillon which decays to a pair of two smaller, {\it constituent} oscillons performing {\it two-bounces}. This means that they collide two times at the intermediate stage of the evolution before they receive enough kinetic energy to separate to infinity.  In Fig. \ref{plot-fractal} such solutions correspond to $A_0$ for which $\phi(0,t)$ passes through two dark lines.

Importantly, in each example, the constituent oscillons perform a certain number of amplitude modulations. These modulations provide a mechanism that explains the chaotic pattern. Namely, if the amplitude decreases the width of the constituent oscillon increases. This is visible as the emergence of augmentation of modulations at the origin, see Fig. \ref{2-b}, lower panel.  As a consequence, this allows to reveal an attractive interaction between the constituent oscillons. Once, the amplitude of the oscillon increases, its width decreases and the oscillon's interaction weakens, which eventually may lead to free oscillons escaping to infinities. 

\begin{figure}
\includegraphics[width=0.85\columnwidth]{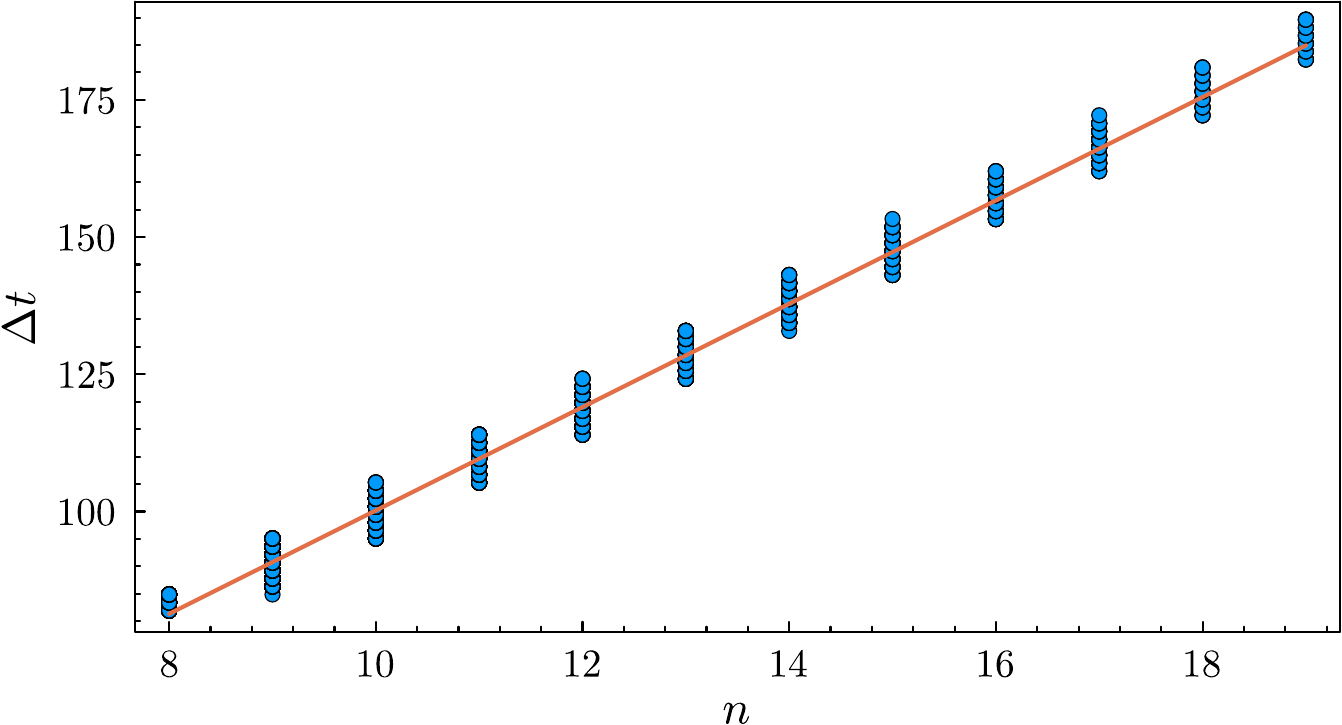}
\caption{Time $\Delta t$ between the first two bounces as a function of the number of modulations. The red line denotes the fit of the linear formula (\ref{fit}).}
\label{window}
\includegraphics[width=0.85\columnwidth]{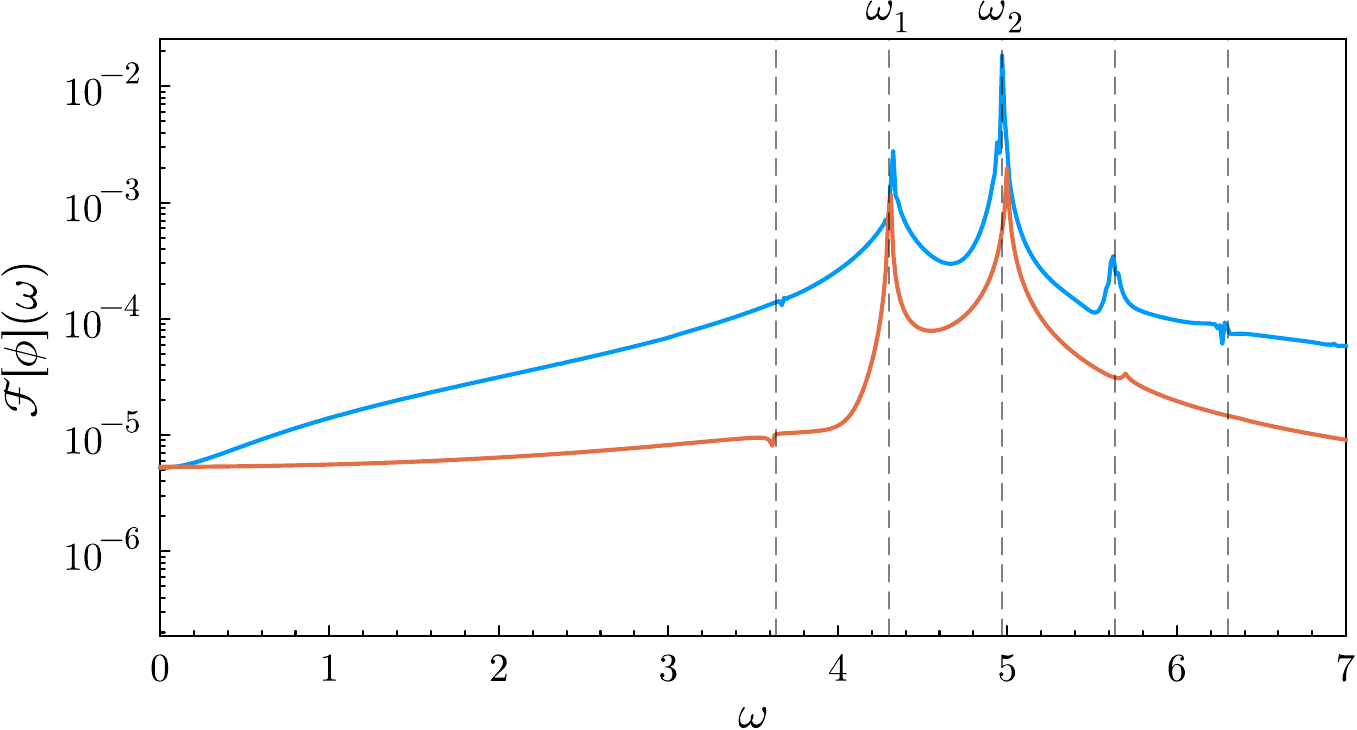}
\caption{Power spectrum. {\it Blue}: $\phi(0,t)$ for a long 2-bounce solution. Here $A_0=0.13745031$ and for $t \in [20,465]$. {\it Orange}: $\phi(2,t)$ for single oscillon generated from a squeezed breather. }
\label{power} 
\end{figure}

In Fig. \ref{window} we plot the time between the first two, long-lasting bounces for $A\in [0.13710, 0.13744]$ as a function of a number $n$ of the amplitude modulations. After fitting a linear expression \be
\Delta t = \frac{2\pi}{\omega_{mod}} n 
\label{fit}
\ee
we find $\omega_{mod}=0.6674 \pm 0.0017$. Hence, the modulations take over the role of the shape mode of a kink \cite{CSW} (or a sphaleron \cite{ACORW}) in triggering the 2-bounce windows. This is confirmed in Fig. \ref{power}, blue curve, which shows the power spectrum of a very long-lasting 2-bounce solution that performs 36 full modulations. Here $A_0=0.13745031$. We identify two main frequencies, $\omega_1\approx 4.30$, which is the frequency of the fundamental oscillations of the oscillon, and $\omega_2\approx 4.97$, which is almost on the mass threshold. Their difference gives the modulations with $\omega_{mod}$. 

To summarize, the excited oscillon can be viewed as a bound state of two constituent oscillons with modulations. Following that, the resonant energy transfer is triggered by two purely oscillon's DoFs. They are the kinetic motion of each constituent oscillon and internal energy stored in the amplitude modulations which play the role of the shape mode in the multi-kink collisions. Therefore, amplitude modulation is not only an interesting characteristic of oscillons but it is a very important factor determining their interactions.

\begin{figure}
\includegraphics[width=1.00\columnwidth]{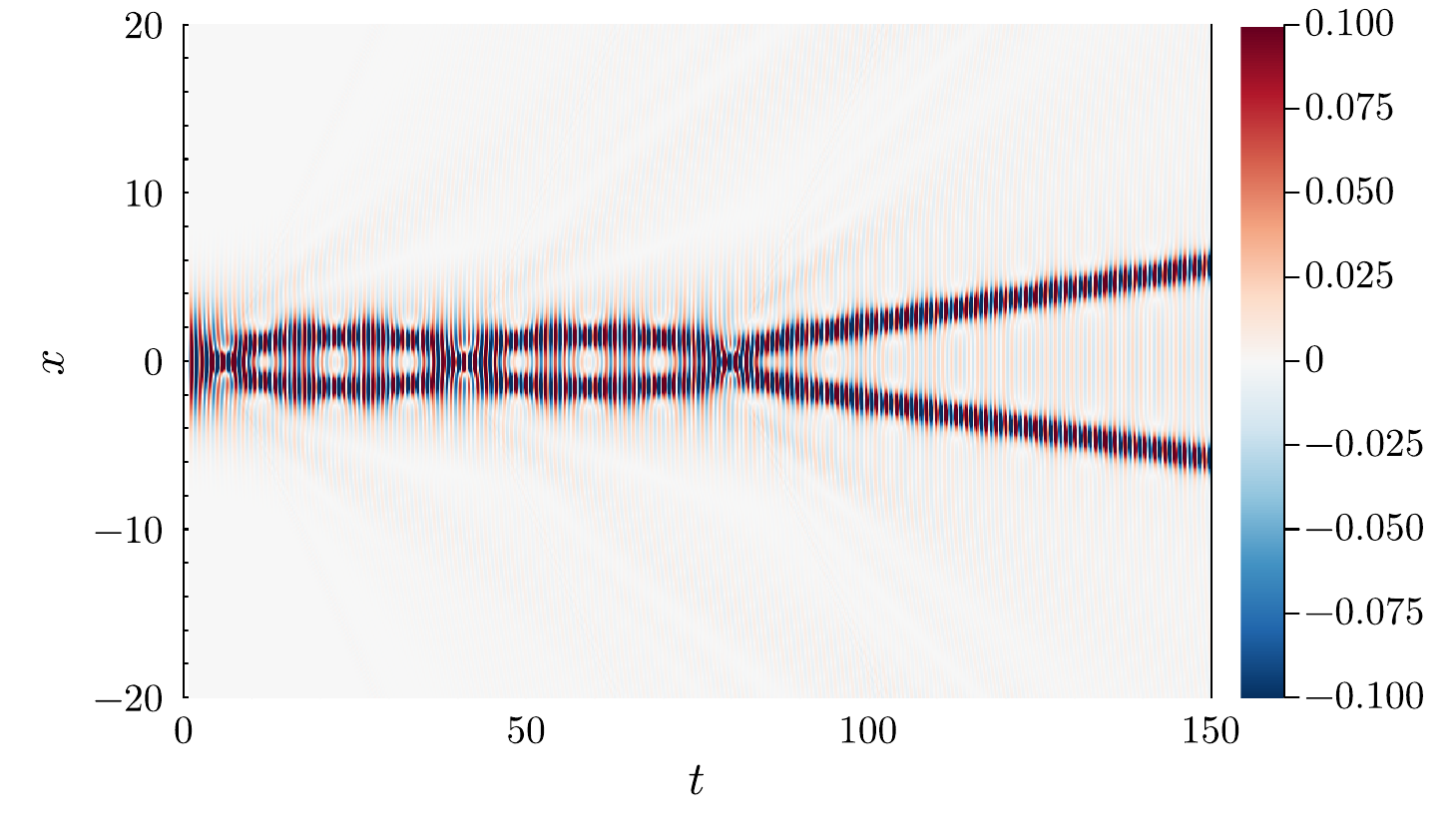}
\caption{An example of 3-bounce of two-oscillon solutions. Here $A_0=0.13356$.}
\label{3-b}
\includegraphics[width=1.00\columnwidth]{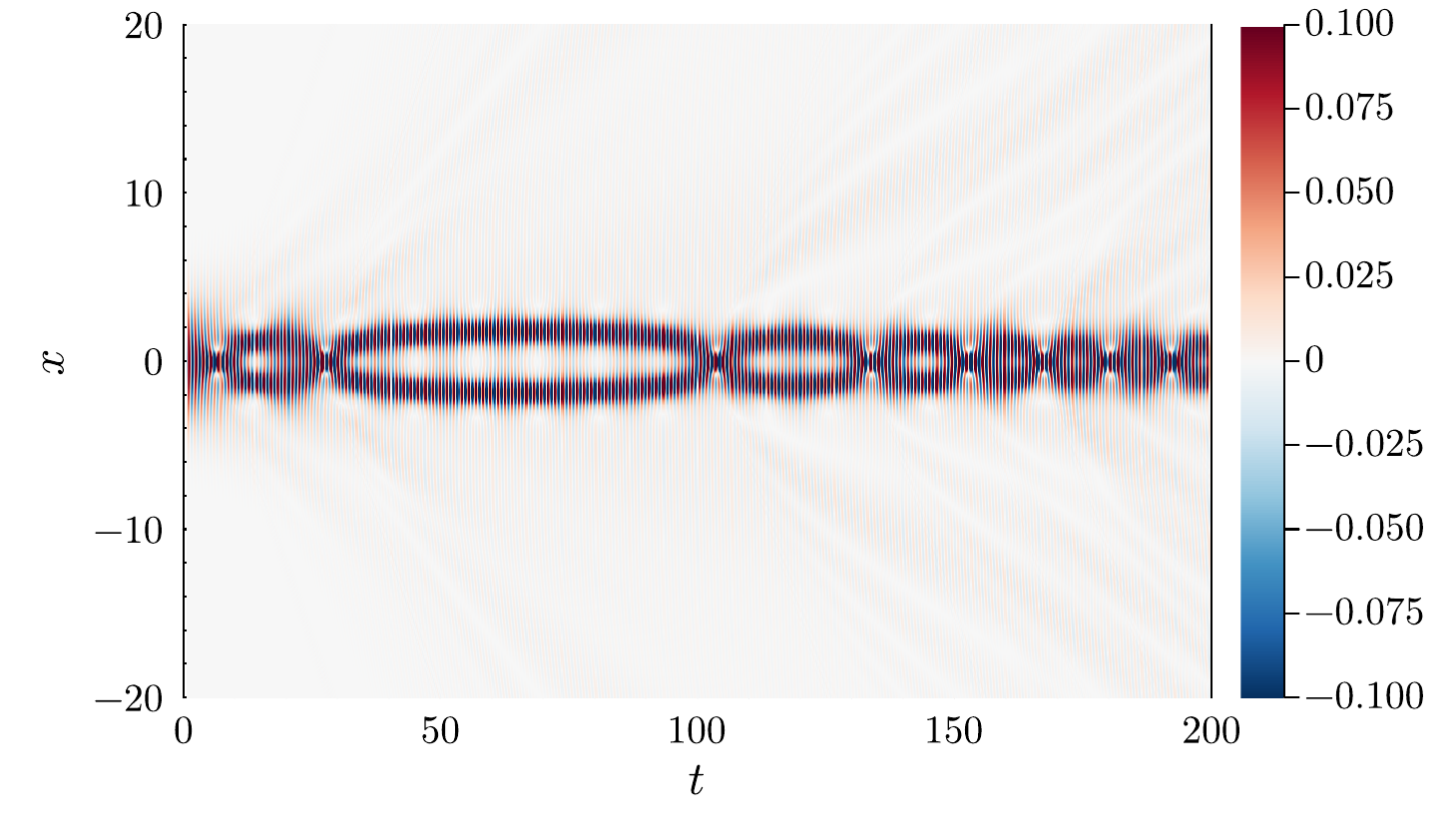}
\caption{An example of the 2-oscillon bion. Here $A_0=0.12$.}
\label{bion}
\end{figure}

We also identified the solutions exhibiting three or more number of bounces. In Fig. \ref{plot-fractal}, they correspond to $A_0$ for which $\phi(0,t)$ goes through a bigger number of dark lines. A particular case with three bounces is shown in Fig. \ref{3-b}.

Finally, we found a solution where the oscillons perform a large number of bounces which eventually lead to a formation of one oscillon at the origin, see Fig. \ref{bion}. Here the attractive force due to the enlargement of the size of the oscillons is never overcome by the kinetic motion of the constituent oscillons. Therefore they never separate. This resembles bion chimneys in kink-antikink collisions.

This simple picture on $n$-bounce windows and ``bion" chimneys is in reality more complicated. Namely, in the decay process, three or more oscillons can be produced. This is not surprising as it is easy to create oscillons with rather small energy (for a bound on the energy see \cite{GS}). In Fig. \ref{plot-fractal} such multi-oscillon solutions are regions with a very dense sequence of horizontal dark lines. For example, we found two satellite, constituent oscillons bouncing around a central one, see the supplementary material.  Again, the modulations of the amplitudes provide attractive interaction that can balance the kinetic motion of the satellite oscillons. 

%%%%%%%%%%%%%%%%%%%%%%%%%%%
\section{\label{sec:origin} Origin of modulations}
%%%%%%%%%%%%%%%%%%%%%%%%%%%

To analyze the constituent oscillons that form the two-oscillon  bound states, responsible for the $n$-bounce windows and bions, and understand the origin of the modulations, it is useful to observe that for small values of the field our potential (\ref{pot}) is very well approximated by the sine-Gordon potential  
\begin{equation}
    U_{sG}(\phi)=\frac{m^2}{6(m^2-1)} \left(1-\cos(\sqrt{6(m^2-1)}\phi)\right).
\end{equation}
The difference appears only at $O(\phi^6)$ order.
Following that, the sine-Gordon breathers can be used as a good approximation for the {\it fundamental} i.e., unexcited, single frequency (small amplitude) oscillons
\begin{equation}
    \phi_B=\frac{4}{\sqrt{6(m^2-1)}}\tan^{-1}\left(\frac{\sqrt{m^2-\omega^2}\cos(\omega t)}{\omega\cosh(\sqrt{m^2-\omega^2}\,x)}\right).
\end{equation}

A weakly excited, constituent oscillon is obtained by a soft squeezing of the breather. Specifically, we assumed its parameters that the oscillon's fundamental (breather) frequency is $\omega_1=4.30$. However, in the effect of the excitement another frequency shows up below the mass threshold, see Fig. \ref{power} orange curve, where we plot the power spectrum of this field at $x=2$. It begins from the mass threshold and then, as the perturbation grows, it slightly decreases to $\omega_2\approx 4.97$. The amplitude modulations are the effect of the interaction of these two frequencies. Again, the frequency of the modulations is $\omega_{mod}=\omega_2-\omega_1\approx 0.667$. These modulations enter the resonant energy transfer. 
\begin{figure}
\includegraphics[width=1.00\columnwidth]{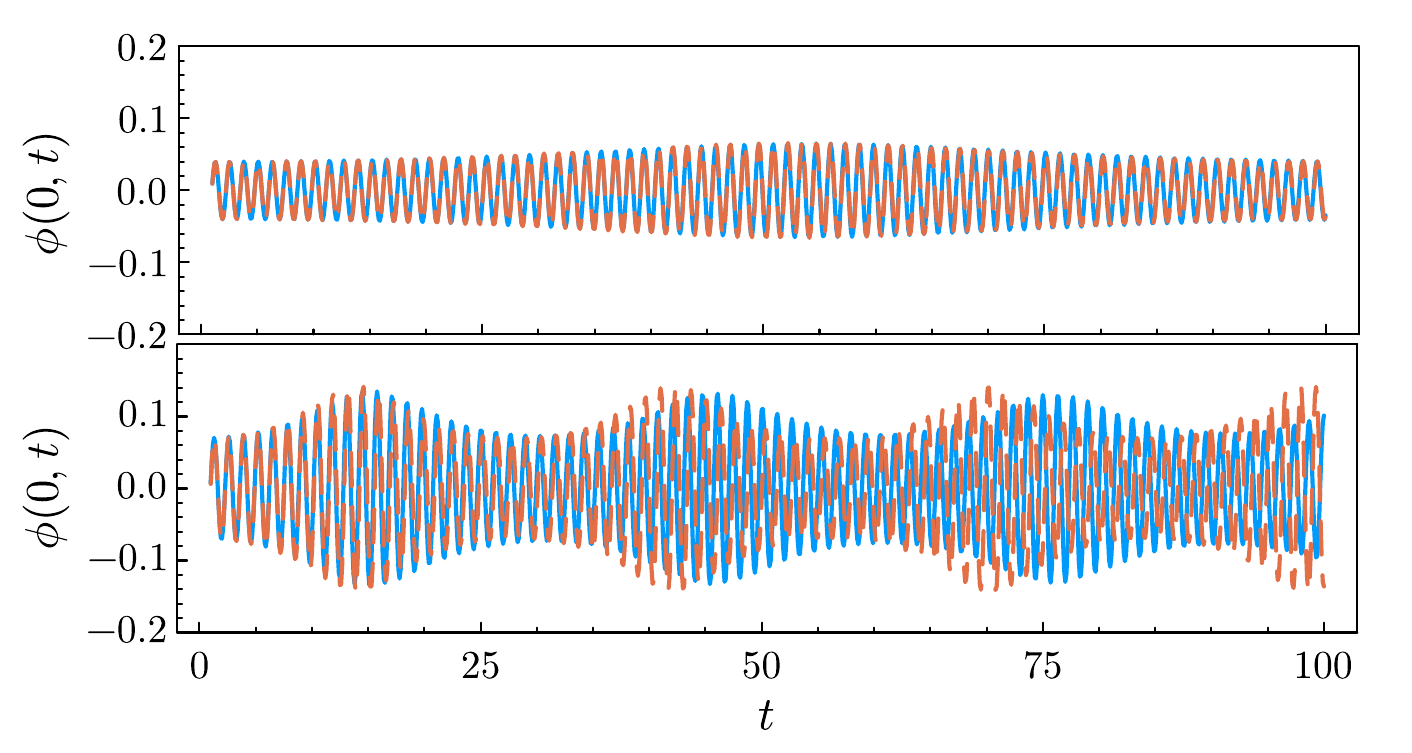}
\caption{Comparison of the evolution of the two-breather solution in sine-Gordon (orange) and our model (blue). Here $\sigma=10$ and $A_0=0.04$ (upper panel), $A_0=0.07$ (lower panel).}
\label{breather}
\end{figure}

Importantly, the appearance of the second frequency suggests that the constituent oscillon with modulations can be viewed as a state of two quasi-breathers. In Fig. \ref{breather} we compare the evolution of the exact two-breather solution in the sine-Gordon theory and in our model. Specifically, the initial configuration is the two-breather $\phi_{BB}(x,t)$ parameterized by $\omega_{1,2}$ such that the Taylor series of $\phi_{BB}(x,0)$ matches Taylor series of $A_0 e^{-x^2/\sigma}$ up tp the second order. This allows to express the parameters $A_0, \sigma$ of Gaussian profile with $\omega_{1,2}$. For amplitudes smaller than $A_0\approx 0.1$  we observe a spectacular agreement. As the amplitude grows (while $\sigma$ is kept fixed) the second oscillon gets excited and the two-breather picture is a bit less accurate. Nonetheless, there is no doubt that the modulations of the amplitude can be attributed to the motion of two quasi-breathers.

%%%%%%%%%%%%%%%%%%%%%%%%%%%
\section{\label{sec:summary}Summary}
%%%%%%%%%%%%%%%%%%%%%%%%%%%
In the current work, we have shown that excited, finite-size oscillons can be viewed as multi-quasi-breather states. Specifically, weakly excited, {\it constituent} oscillons are solutions composed of two {\it fundamental} oscillons identified with breathers. This immediately results in the appearance of modulations of the amplitude - a crucial oscillon observable whose origin has been extensively investigated, see e.g., \cite{MR, J}. 

Next, largely excited oscillons can be treated as four (or even more complicated) quasi-breather states, with two pairs of quasi-breathers forming two constituent oscillons. This leads to the appearance of a fractal pattern in the decay of strongly excited oscillons, which can be explained by the resonant energy transfer mechanism involving the kinetic motion of the constituent, weakly excited oscillons, and their amplitude modulations. A net effect of these two DoFs leads to chaotic behavior in the final state formation which resembles the famous fractal structure in kink-antikink collisions in non-intergrable models. Importantly, all DoFs are provided only by the oscillons themselves as there are no other topological or non-topological solutions that could be a source of internal vibrations. 

We underline that even though we used a specific potential similar fractal patterns should exist in other models, e.g., \cite{Z}. This is based on the observation that bounces of oscillons have previously been reported \cite{BPS-S, Simas}, although never appreciated or studied properly. Of course, the existence of localized static solutions like kinks or sphalerons may add new DoF to the decay process \cite{SR}. 

It would be interesting to investigate any possible relation between a multi-quasi-breather picture of excited oscillons and their decomposition into Derrick modes \cite{J}. 

It is also tempting to conjecture that bounces of (1+1)-dimensional oscillons may be related to an observation that decay of (3+1)-dimensional oscillons is preceded by a regime of more pronounced amplitude modulations \cite{H, GGB}. Furthermore, probably multipolar structures observed in an excited oscillon may also be associated with the motion of unexcited oscillons \cite{WXZ}. 

%%%%%%%%%%%%%%%%%%%%%%%%%%%
\section*{Acknowledgements}
%%%%%%%%%%%%%%%%%%%%%%%%%%%%

KS, TR and AW were supported by the Polish National Science Center, 
grant NCN 2019/35/B/ST2/00059 and the Priority Research Area under the
program Excellence Initiative—Research University at
the Jagiellonian University in Kraków. FB acknowledges the institutional support of the Research Centre for Theoretical Physics and Astrophysics, Institute of Physics, Silesian University in Opava. AW acknowledges support of the Spanish Ministerio de Ciencia e Innovacion (MCIN) with funding from the European Union NextGenerationEU (PRTRC17.I1) and the Consejeria de Educacion, Junta de Castilla y Leon, through QCAYLE project, as well as grant PID2020-113406GB-I00 MTM funded by MCIN/AEI/10.13039/501100011033. 

\vspace*{0.5cm}

%%%%%%%%%%%%%%%%%%%%%%%%%%%
\section*{Supplementary material} 
%%%%%%%%%%%%%%%%%%%%%%%%%%%%

 %%%%%%%%%%%%%%%%%%%%%%%%%%%
\subsection*{A. Evolution of various initial data}
%%%%%%%%%%%%%%%%%%%%%%%%%%%%
\begin{figure*}
\center
\includegraphics[width=2.0\columnwidth]{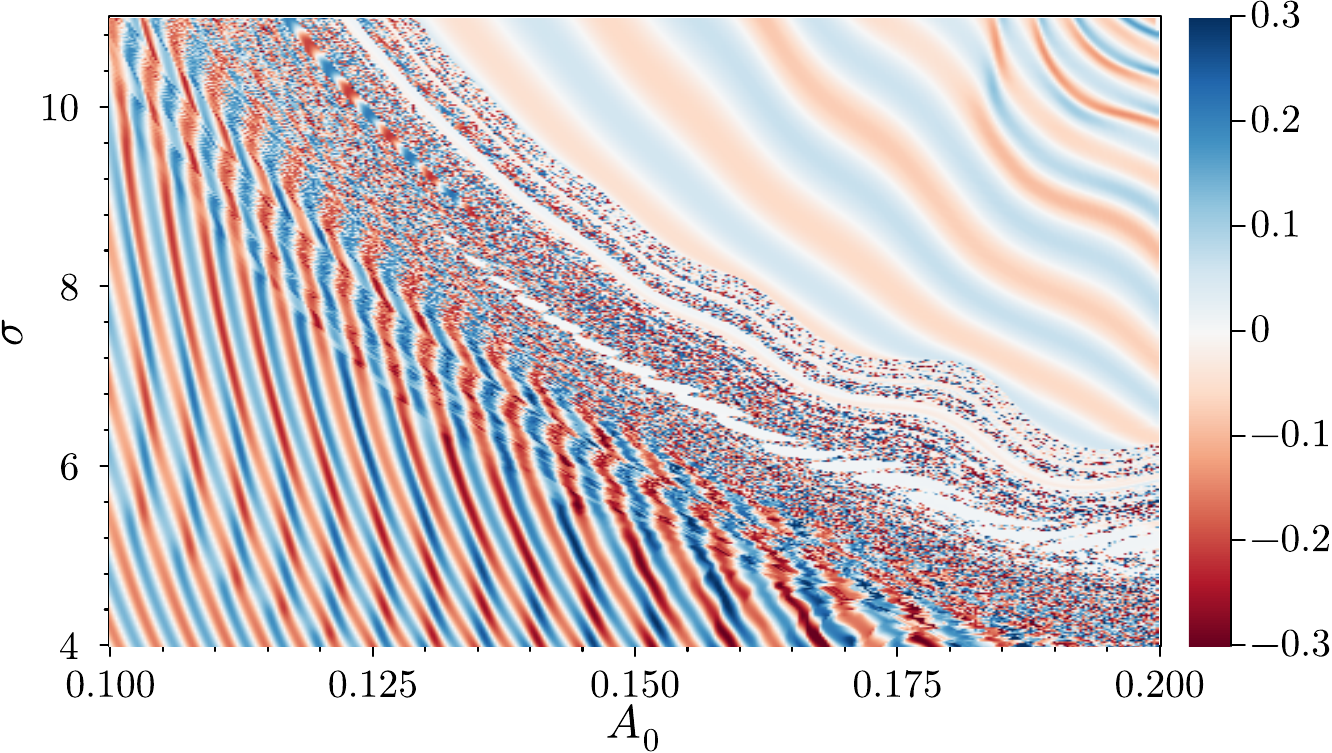}
\caption{The value of the field at the origin normalized to the initial pulse amplitude $A_0$, $\phi(0,T)/A_0$, at large time $T=200$.}
\label{multi-plot}
\end{figure*}

It is instructive to analyze the evolution of the initial Gaussian profile for a large set of parameters. Specifically, we considered $\sigma \in [4,11]$ and $A_0\in [0.1, 0.2].$ The fractal structure of interchanging multi-bounce windows and bion-like chimneys is a genuine effect existing for any $\sigma$. 

In Fig. \ref{multi-plot} we present a scan displaying the value of the field at the origin at $T=200$, divided by the initial amplitude, i.e. $\phi(0,T)/A_0$. We find a complicated structure where white regions (no oscillon at the origin at $T=200$) are chaotically immersed between very thin multi-color regions (oscillons at the origin at $T=200$). If two oscillons are produced, these regimes may be identified with bounce windows and "bion" chimneys. However, the existence of multi-oscillon solutions makes this simple identification more involved. For example, we found solutions where two oscillons are bouncing around a third one, which remains localized at the origin. This obviously amounts to a multi-color region in the scan. 

We see that as $\sigma$ and $A_0$ get smaller the initial configuration produces mainly a weakly excited oscillon seen as wider multi-color regions.

\begin{figure*}
\center
\includegraphics[width=1.0\columnwidth]{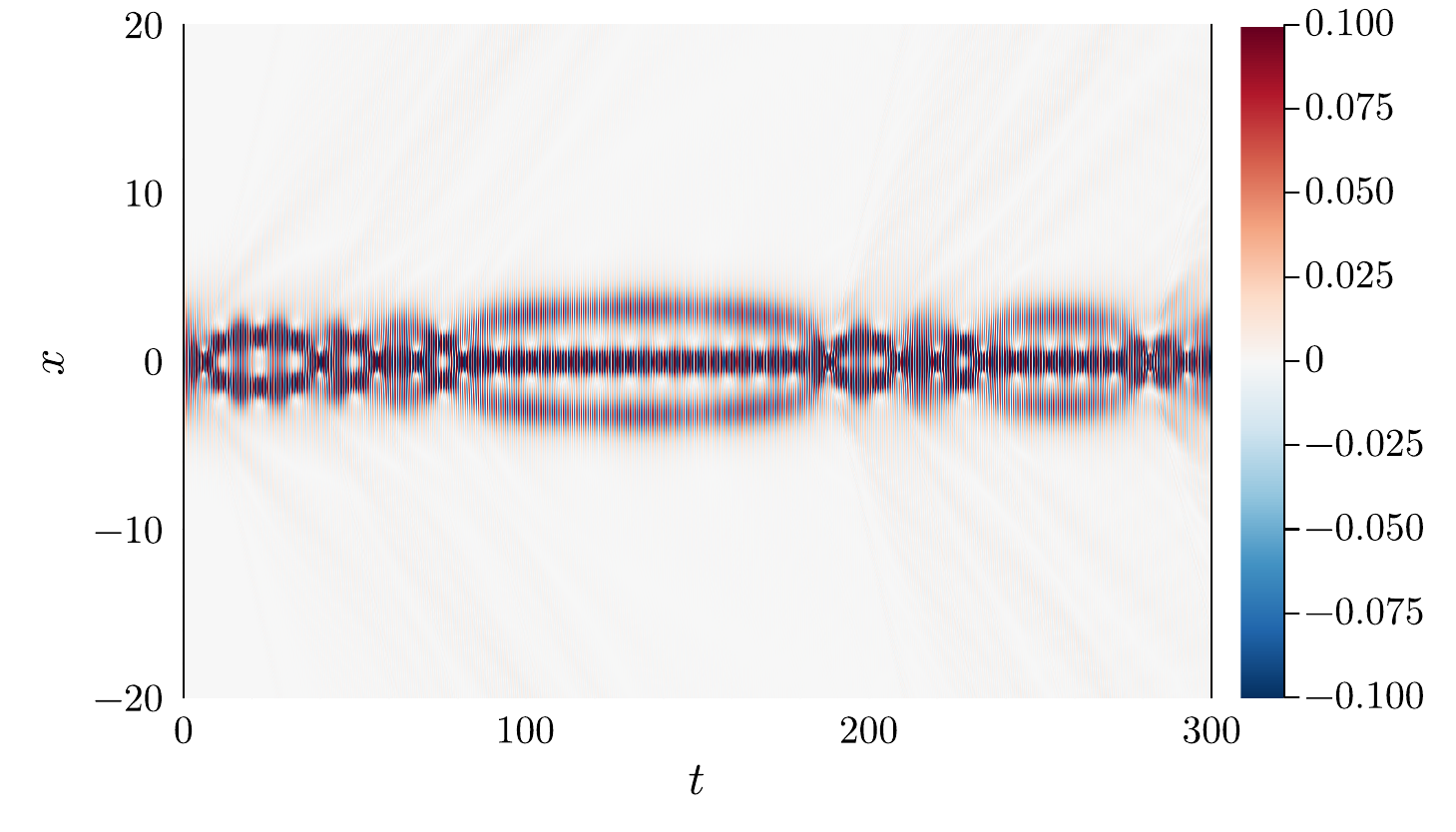}
\includegraphics[width=1.0\columnwidth]{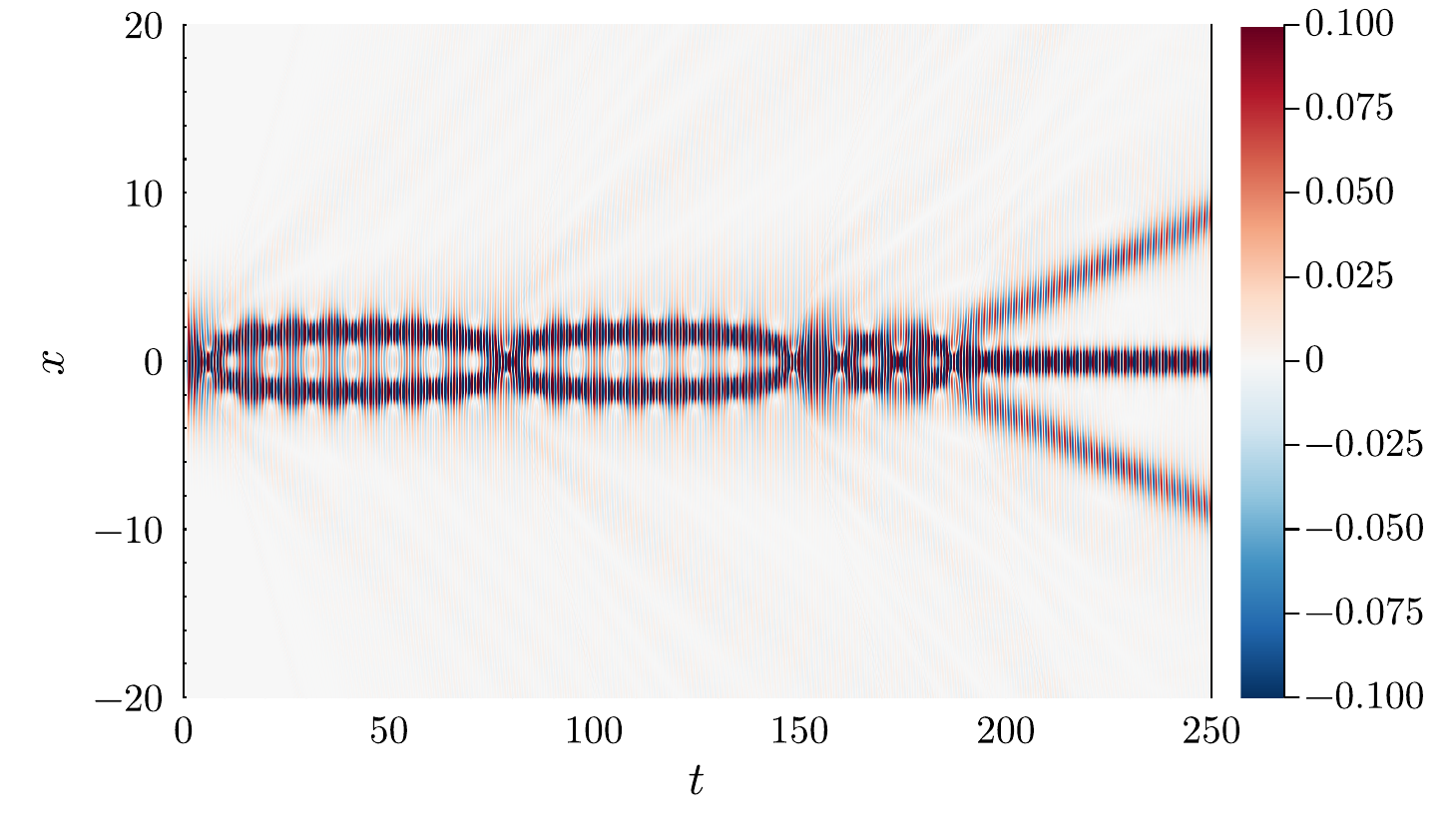}
\caption{Examples of three-oscillon solutions. $\phi(x,t)$ for $\sigma=10$,  $A_0=0.13312$ (left panel) and $A_0=0.136933$ (right panel).}
\label{3-oscillon}
\end{figure*}
 %%%%%%%%%%%%%%%%%%%%%%%%%%%
\subsection*{B. Multi oscillon states}
%%%%%%%%%%%%%%%%%%%%%%%%%%%%
As we remarked before, an initial Gaussian profile can easily lead to more than two oscillon states. This is because there is no topological bound on the energy of the oscillon. Therefore, they are quite light quasi-particle objects. Of course, such multi-oscillon configurations usually appear for bigger initial amplitude $A_0$. However, we observed some deviations from this naive expectation. Indeed, increasing the energy of the initial pulse (initial amplitude) may lead to the appearance of a smaller number of oscillons. A detailed study of this issue goes beyond the scope of the current work. 

In Fig. \ref{3-oscillon} we present examples of a solution involving three oscillons. In the left panel, we show a bion-type case where the constituent oscillon at the origin is accompanied by two satellite oscillons. In the upper right panel, we have a bounce-type solution where in the final state we find three oscillons escaping from each other.

If the initial amplitude grows even more complicated solutions can be obtained. 

%%%%%%%%%%%%%%%%%%%%%%%%%%%
\subsection*{C. Two breather solution}
%%%%%%%%%%%%%%%%%%%%%%%%%%%%

Here we give an exact expression for the two-breather solution in the sine-Gordon model
\begin{widetext}
\be 
\Phi_{BB}(x,t)=4 \tan^{-1} 
 \left( \frac{2a_1 \cosh\left( \sqrt{1-\omega_2^2} x \right) \cos (\omega_1 t) +2a_2 \cosh\left( \sqrt{1-\omega_1^2} x \cos (\omega_2 t) \right)}{\cosh  \left( \Omega_+ x \right) + A^{-1}_- A^{-1}_+ \cosh  \left( \Omega_- x \right) + a_1a_2 \left(A_- \cos((\omega_1+\omega_2)t) +A_+ \cos((\omega_1-\omega_2)t)  \right) } \right)
\ee
\end{widetext}
where $\omega_{1,2}=\sin \alpha_{1,2}$ and
\bea
\Omega_{\pm}&=&\sqrt{1-\omega_1^2} \pm \sqrt{1-\omega_2^2} \nonumber \\
A_\pm &=& \tan^2 \left( \frac{\alpha_1 \pm \alpha_2}{2} \right), \nonumber \\
a_{1,2} &=& \frac{1}{\tan(\alpha_{1,2})\sqrt{A_+ A_-}}.
\eea
In our analysis, we used a rescaled version of it
\be
\phi_{BB}(x,t) = \frac{1}{\sqrt{6(m^2-1)}} \Phi_{BB}(mx,mt).
\ee


\newpage
\begin{thebibliography}{99}
\bibitem{G} M. Gleiser, Pseudostable bubbles, Phys. Rev. D49 (1994) 2978.

\bibitem{BM} I. L. Bogolyubsky and V. G. Makhankov, On the pulsed soliton lifetime in two classical relativistic theory models, JETP Lett. 24 (1976) 12.

\bibitem{GS} M. Gleiser and D. Sicilia, A General Theory of Oscillon Dynamics, Phys. Rev. D 80 (2009) 125037.

\bibitem{I} M. Ibe, M. Kawasaki, W. Nakano, E. Sonomoto, Decay of I-ball/Oscillon in Classical Field Theory, JHEP 04 (2019) 030.

\bibitem{Z} H.-Y. Zhang, M. A. Amin, E. J. Copeland, P. M. Saffin, K. D. Lozanov, Classical Decay Rates of Oscillons, JCAP 07 (2020) 055. 

\bibitem{CGM} E. J. Copeland, M. Gleiser, and H. R. Muller, Oscillons: Resonant configurations during bubble collapse, Phys. Rev. D 52 (1995) 1920.

\bibitem{G-cosm} M. Gleiser, Oscillons in scalar field theories: Applications in higher dimensions and inflation, Int. J. Mod. Phys. D 16 (2007) 219.

\bibitem{Amin-1} M. A. Amin and D. Shirokoff, Flat-top oscillons in an expanding universe, Phys. Rev. D81 (2010) 085045.

\bibitem{A} M. A. Amin, R. Easther, H. Finkel, R. Flauger and M. P. Hertzberg, Oscillons After Inflation, Phys. Rev. Lett. 108 (2012) 241302.

\bibitem{FMPW} R. Flauger, L. McAllister, E. Pajer, A. Westphal, G. Xu Oscillations in the CMB from Axion Monodromy Inflation, JCAP 06 (2010) 009. 

\bibitem{AE} M. A. Amin, R. Easther, Inflaton Fragmentation and Oscillon Formation in Three Dimensions, JCAP 12 (2010) 001. 

\bibitem{LT} K. D. Lozanov, V. Takhistov, Enhanced Gravitational Waves from Inflaton Oscillons,  Phys. Rev. Lett. 130 (2023) 181002. 



\bibitem{G-weak} N. Graham, An Electroweak Oscillon, Phys. Rev. Lett. 98 (2007) 101801, [Erratum-ibid. 98, 189904 (2007)].

 \bibitem{Fod} G. Fodor, P. Forgacs, Z. Horvath, and A. Lukacs, Small amplitude quasibreathers and oscillons, Phys. Rev. D 78 (2008) 025003.

 \bibitem{DRSW} P. Dorey, Y. Shnir, T. Romanczukiewicz, and A. Wereszczynski, Oscillons in gapless theories, arXiv:2312.05308.
 
\bibitem{MR} N. Manton and T. Romanczukiewicz, The Simplest Oscillon and its Sphaleron, Phys. Rev. D 107 (2023) 085012.

\bibitem{OQRW} K. Oles, J. Queiruga, T. Romanczukiewicz, A. Wereszczynski, Sphaleron without shape mode and its oscillon, Phys. Lett. B 847 (2023) 138300.

\bibitem{sug}  T. Sugiyama, Kink-antikink collisions in the two-dimensional $\phi^4$ model, Prog. Theor. Phys. 61 (1979) 1550.

\bibitem{CSW}  D. K. Campbell, J. F. Schonfeld, and C. A. Wingate,
Resonance structure in kink-antikink interactions in $\phi^4$
theory, Physica D 9 (1983) 1.

\bibitem{MORW} N. S. Manton, K. Oles, T. Romanczukiewicz, and A.
Wereszczynski, Collective coordinate model of kink-antikink collisions in $\phi^4$ theory, Phys. Rev. Lett. 127 (2021) 071601.

\bibitem{ACORW} C. Adam, D. Ciurla, K. Oles, T. Romanczukiewicz, and A. Wereszczynski, Sphalerons and resonance phenomenon in
kink-antikink collisions, Phys. Rev. D 104 (2021) 105022.

\bibitem{J} S. Navarro-Obregon, L. M. Nieto, and J. M. Queiruga, Inclusion of radiation in the CCM approach of the $\phi^4$ model, Phys. Rev. E 108 (2023) 044216. 

\bibitem{BPS-S} A. Alonso-Izquierdo, S. Navarro-Obregon, K. Oles, J. Queiruga, T. Romanczukiewicz, and A. Wereszczynski, Semi-Bogomol’nyi-Prasad-Sommerfield sphaleron and its dynamics, Phys. Rev. E 108 (2023) 064208.

\bibitem{Simas} F. C. Lima, F. C. Simas, K. Z. Nobrega, and A. R. Gomes,
Scattering of metastable lumps in a model with a false vacuum,
Phys. Lett. B 822 (2021) 136707.

\bibitem{SR} Y. Shnir, and T. Romanczukiewicz, Oscillon resonances and creation of kinks in particle collisions, Phys. Rev. Lett. 105 (2010) 081601. 


\bibitem{H} E. P. Honda, M. W. Choptuik, Fine structure of oscillons in the spherically symmetric $\phi^4$ Klein-Gordon model, Phys. Rev. D 65 (2002) 084037.

\bibitem{GGB} J. T. Gálvez Ghersi, J. N. Braden, Dimensional deformation of sine-Gordon breathers into oscillons, Phys. Rev. D 108 (2023) 096017. 

\bibitem{WXZ} Y.-J. Wang, Q.-X. Xie, and S.-Y. Zhou, Excited oscillons: Cascading levels and higher multipoles, Phys. Rev. D 108 (2023) 025006.


\end{thebibliography}
\end{document}